# Observation of the hybridization gap and Fano resonance in the Kondo lattice URu$_2$Si$_2$

W. K. Park[1,*], P. H. Tobash[2], F. Ronning[2], E. D. Bauer[2], J. L. Sarrao[2], J. D. Thompson[2], and L. H. Greene[1]

[1]*Department of Physics and the Frederick Seitz Material Research Laboratory, University of Illinois at Urbana-Champaign, Urbana, Illinois 61801, USA*

[2]*Los Alamos National Laboratory, Los Alamos, New Mexico 87545, USA*

Abstract

The nature of the second order phase transition that occurs in URu$_2$Si$_2$ at 17.5 K remains puzzling despite intensive research over the past two and half decades. A key question emerging in the field is whether a hybridization gap between the renormalized bands can be identified as the long-sought 'hidden' order parameter. We report on the measurement of a hybridization gap in URu$_2$Si$_2$ employing a spectroscopic technique based on quasiparticle scattering across a ballistic metallic junction. The differential conductance exhibits an asymmetric double-peak structure, a clear signature for a Fano resonance in a Kondo lattice. The extracted hybridization gap opens well above the transition temperature, indicating that it is not the hidden order parameter. Our results put stringent constraints on the origin of the hidden order transition in URu$_2$Si$_2$ and demonstrate that quasiparticle scattering spectroscopy can probe the band renormalizations in a Kondo lattice via detection of a novel type of Fano resonance.

*Corresponding author: wkpark@illinois.edu

The 5$f$ orbital-based heavy electron system URu$_2$Si$_2$ has long puzzled researchers due to its enigmatic transition at $T_{HO}$ = 17.5 K into the hidden order (HO) [1-4]. Despite numerous reports of gap-like behaviors [1-3,5], the exact order parameter remains unknown [4,6-9]. Static antiferromagnetism [1-3] is ruled out because the measured magnetic moment is too small to account for the large entropy loss [10] and has been shown to be extrinsic [10]. Under pressure, the HO undergoes a first order transition into an antiferromagnetic (AF) state [10-12], and can be resurrected by magnetic field [13]. Inelastic neutron scattering (INS) has established two magnetic excitations [14-17]: $\mathbf{Q_0}$ = (1, 0, 0), $E_0$ = 1.7- 2 meV; $\mathbf{Q_1}$ = (1±0.4, 0, 0), $E_1$ = 4 – 5.7 meV. It has become evident that identifying the origin of the $\mathbf{Q_0}$ resonance, a unique feature of the HO, is critical [16]. Differentiating the consequences of the HO transition from its origin is also crucial, as demonstrated here.

Quasiparticle (QP) probes measuring tunneling and scattering conductance can provide direct electronic structure information. Recent investigations for a Kondo lattice, experimental [18-22] and theoretical [8,23-27], have brought new perspectives on the HO problem. A key question is whether a hybridization gap between the renormalized bands can be identified as the HO parameter [8]. In this Letter, we report spectroscopic measurements of a hybridization gap in URu$_2$Si$_2$ using quasiparticle scattering spectroscopy (QPS) or point-contact spectroscopy [19,28]. Our conductance spectra clearly exhibit characteristic features for a Fano resonance in a Kondo lattice including a distinct asymmetric double-peak structure. Analysis based on a recent theory [23] allows us to extract the hybridization gap: This gap opens well above $T_{HO}$, indicating it is not the HO parameter.

Tunneling in a single Kondo adatom has been extensively investigated using a scanning tunneling microscope (STM) [29,30] and well accounted for by the generic Fano resonance [31] formula: $dI/dV]_{KI} \propto (q_F + E')^2 / (1 + E'^2)$, where $E' \equiv (eV - \varepsilon_0)/(W/2)$ with $\varepsilon_0$ and $W$ being the resonance energy and full width at half maximum, respectively. As demonstrated in Fig. 1a, the Fano factor, $q_F \equiv A/B$ ($A$, tunneling probability into a localized orbital; $B$, into the conduction band), is a key parameter governing the conductance shape. According to the Kondo lattice model, the fate of localized moments is determined by the competition between the Kondo coupling and the Ruderman-Kittel-Kasuya-Yosida (RKKY) interaction [32]. If the former is predominant, a coherent heavy electron liquid emerges, whereas antiferromagnetism is the ground state if the RKKY interaction is stronger. Fermi surface (FS) topology plays important roles not only in itinerant magnetism such as a spin-density wave induced by FS nesting [33] but also in mediating the RKKY interaction between local moments [34]. The periodic Anderson model, in a mean-field approximation considering on-site coulomb interaction, gives two renormalized hybridized bands [35]: $E_{k\pm} = \frac{1}{2}\{\varepsilon_k + \lambda \pm \sqrt{(\varepsilon_k - \lambda)^2 + 4V^2}\}$. Here, $\lambda$ is the renormalized $f$-level and $V = z^{1/2}V_0$ is the renormalized hybridization matrix amplitude with $z = 1-n_f$ ($n_f$: $f$-level occupancy). As shown in Figs. 1b & 1c, a hybridization gap opens: a direct gap of $2V$ in $\bm{k}$-space and an indirect gap in the density of states (DOS) given as $\Delta_{\text{hyb}} = 2V^2/D$ ($2D$: conduction bandwidth). Based on this hybridization picture plus cotunneling, the differential tunneling conductance in a Kondo lattice was derived [23]:

$$\left.\frac{dI}{dV}\right|_{FR} \propto \text{Im}\, \tilde{G}_\psi^{KL}(eV); \quad \tilde{G}_\psi^{KL}(eV) = \left(1 + \frac{q_F W}{eV - \lambda}\right)^2 \ln\left[\frac{eV + D_1 - \frac{V^2}{eV - \lambda}}{eV - D_2 - \frac{V^2}{eV - \lambda}}\right] + \frac{2D/t_c^2}{eV - \lambda},$$

where $-D_1$ and $D_2$ are the lower and upper conduction band edges, respectively. $q_F = t_f V/t_c W$, where $t_f$ and $t_c$ are the tunneling matrix amplitudes for the $f$-orbital and the conduction band, respectively [23]. As shown in Fig. 1d, for an intermediate $q_F$, an *asymmetric double-peak structure* is notable: The hallmark for a Kondo lattice, distinct from the single impurity case.

Single crystalline URu$_2$Si$_2$ and U(Ru$_{0.985}$Rh$_{0.015}$)$_2$Si$_2$ are grown by the Czrochralski method and oriented using a back-Laue CCD camera. The *ab*-plane resistivity and the specific heat of Fig. 2 show our crystals exhibit distinct bulk HO and superconducting transitions. As-grown or cleaved crystals with mirror-like surfaces normal to the *c* axis are used in QPS. Ballistic metallic junctions are formed at low temperature using an electrochemically polished gold tip and differential micrometer [18,19]. Junctions are formed on different spots *in situ* as resistance and pressure are controlled. Differential conductance is measured with a lock-in technique as a function of temperature and magnetic field.

Figures 3a & 3b display a series of conductance curves for URu$_2$Si$_2$ and U(Ru$_{0.985}$Rh$_{0.015}$)$_2$Si$_2$. A systematic evolution in the shape is clearly noticeable with a distinct double-peak structure appearing in curves 3 – 5, from which we conjecture on two parallel channels, one dominating the background and the other the asymmetric double-peak structure. We focus on the latter, leaving the background shape for future investigation. Andreev scattering [18,19] is ruled out since $T \gg T_c$ (= 1.4 K). So is an AF gap [36,37] (also, Refs. in [38]) excluded [10]. To elucidate its origin, we further note the positive-bias peak is always stronger and the conductance minimum occurs at a negative bias (typically, −0.5 ~ −3 mV at $T \ll T_{HO}$), supporting the Fano resonance origin (Fig. 1d [23], S1 in [38]).

For a quantitative analysis, we start by considering strongly energy-dependent QP scattering into the renormalized heavy bands: The larger the DOS, the higher the transition rate. QPs passing through two channels, the heavy and the conduction band, interfere to produce a Fano resonance. In recent STM studies on Kondo adatoms [39], a single-impurity Fano resonance was observed in the metallic contact regime as well as in the tunneling regime, indicating that a similar quantum interference occurs in both regimes. Therefore, we conjecture that the afore-described Kondo lattice tunneling theory [23] can account for the characteristic features in our QPS data. The same Fano physics manifests in both QP tunneling and scattering with the conductance shape dictated by the universal parameter, $q_F$. Thus, our model formula is:

$$G(V) \equiv \frac{dI}{dV} = \left.\frac{dI}{dV}\right|_{FR} + \omega \cdot \left.\frac{dI}{dV}\right|_{bg},$$

where the first term is the Fano resonance conductance and the second term accounts for the background shape with $\omega$ as a weighting factor. Figures 3c-f show typical conductance curves for URu$_2$Si$_2$ and best fits obtained with a parabolic background and an energy-dependent QP broadening parameter [26]. Our model captures major conductance features accurately (S2 in [38]). The hybridization gap is extracted from the fitting parameters using the relation $\Delta_{hyb} = 2V^2/D$. It ranges from 11 - 14 meV with an average of 13 meV. The renormalized hybridization strength $V$ = 39 - 45 meV and the Fano parameter $q_F$ = 9 - 13. These values are reproducibly observed in many more conductance curves [38]. For U(Ru$_{0.985}$Rh$_{0.015}$)$_2$Si$_2$ with $T_{HO}$ = 12.8 K (Fig. 2a), $\Delta_{hyb} \approx$ 10 meV, implying some correlation (proportionality) with $T_{HO}$.

The relation between the HO transition and the hybridization process [8] is addressed in Fig. 4a, showing the temperature-dependent conductance spectra with best fits (S2 in [38]). The split peaks persist across $T_{HO}$, disappearing at a much higher temperature. The temperature dependence of $\Delta_{hyb}$ is plotted in Fig. 4b. Note the hybridization gap reproducibly opens at $T_{hyb}$ ~ 27 K (S3 in [38]), well above $T_{HO}$, establishing that the gap opening well precedes the HO transition. Of the published QPS data, we note that the sharper the low-temperature gap structure is, the higher the gap opening temperature is observed (S3 in [38]). The renormalized $f$-level, $\lambda$, appears to cross the chemical potential, $\mu$, at $T \approx T_{HO}$ (Fig. 4b). The spectral peak in a recent angle-resolved photoemission spectroscopy (ARPES) study [40] shows a similar behavior, which can be understood by assuming a broadening-induced merging of the hybridization-gap peaks into a single Kondo resonance peak [41]. Considering $\varepsilon_0 = W/2 \cdot tan[(1-n_f)\pi/2]$, the sign change in $\lambda$ ($\varepsilon_0$) may signify a $f$-level occupancy change accompanying the HO transition; Further investigation is necessary to elucidate its physical meaning more clearly. The normalized zero-bias conductance (NZBC) also reveals non-trivial temperature dependence as plotted in Fig. 4c, a broad maximum around $T_{HO}$ [37]. A hallmark of a QPS junction being in the thermal (non-spectroscopic) regime is that $G(V)$ (also, ZBC($T$)) strongly resembles the bulk conductivity [19,28,38]. That our data do not exhibit such a behavior indicates the junctions are well within the spectroscopic limit [19,28,38]. To account for this temperature dependence, first note the NZBC would be proportional to the DOS at $\mu$ for tunneling into the heavy band only, thus, to the electronic specific heat coefficient ($C_e/T$) and effective mass. Indeed, $C_e/T$ is found to show qualitatively similar temperature dependence [42]. A large contribution from the heavy band (large $q_F \sim 10$) as well as the ballistic nature enables us to observe such a behavior (S4 in [38]).

Our earlier QPS studies on CeCoIn$_5$ [18-20] have shown a single impurity-like Fano line shape, contrary to URu$_2$Si$_2$. Considering $\Delta_{hyb} = 2V^2/D$ and $V = V_0(1 - n_f)^{1/2}$, we conjecture this discrepancy may arise from their different distances from the Kondo regime ($n_f \sim 1$) [43]: $\Delta_{hyb}$ would become smaller away from it ($n_f < 1$), rendering the peaks more susceptible to merging. This agrees with CeCoIn$_5$ being usually considered closer to the Kondo limit than URu$_2$Si$_2$ [44]. The distinct double-peak structure seen in our data implies the broadening effect, suggested to arise from intrinsic correlation [26], lattice disorder [23], and broken translation invariance [26], is not dominant in URu$_2$Si$_2$. In recent STM studies on URu$_2$Si$_2$ [21,22], a single impurity Fano line shape is observed with $q_F < 2$, implying the tunneling probability into the heavy band is much lower than in our QPS, which can account for the line shape (S4 in [38]). Other disparate STM observations are [21,22]: i) gap opening at 16 - 17 K; ii) gap size of ~ 8 meV; iii) fine structures at low bias and temperature. To account for these discrepancies, one may consider surface effects, i.e., possible modifications in the hybridization [45] due to reduced near-neighbor coordination [30]. QPS (Fig. 1d inset) in the ballistic regime [19,28] probes scattering over the electronic mean free path, well beyond the surface. Thus, QPS is more likely to detect the bulk hybridized bands, as manifested by higher $T_{hyb}$ and robust double-peak structure as predicted [23]. A recent optical spectroscopy, known as a bulk probe, has reported similar $\Delta_{hyb}$ and $T_{hyb}$ values [46].

We now address the widely varying gap values extracted from other measurements [1-3,5] by focusing on resistivity (S5 in [38]). Despite no evidence for static magnetism, resistivity is frequently analyzed considering scattering off gapped magnetic excitations ($\rho_m$): $\rho = \rho_0 + AT^2 + \rho_m$. Furthermore, nearly all reports have adopted a formula for ferromagnetic (FM) excitations [47]: $\rho_{FM} = BT\Delta[1 + 2T/\Delta]e^{-\Delta/T}$, despite the close proximity to an AF order; and $\rho_{AF}$ takes a quite different form due to linear, not quadratic, dispersion. Two known approximate formulae are: $\rho_{AF1} = B\Delta^5[(T/\Delta)^5/5 + (T/\Delta)^4 +$ $5/3 \cdot (T/\Delta)^5]e^{-\Delta/T}$ [48] and $\rho_{AF2} = B\Delta^2\sqrt{(T/\Delta)}[1 + 2/3 \cdot (T/\Delta) + 2/15 \cdot (T/\Delta)^2]e^{-\Delta/T}$ [49]. In order to include the transition region (Fig. 2a), we adopt a generic $T$-dependent $\Delta(T) = \Delta_0 tanh[\alpha\sqrt{(T_{HO}/T-1)}]$. The formulae based on FM excitations [47] and FS gapping [17] give diverging fits as $T$ approaches $T_{HO}$, whereas the two AF formulae produce reasonably good fits including the transition region, with $\Delta_0 \sim 5$ meV and $\alpha = 1.7$, as shown in Fig. 2a [50]. Our analysis extended to other published data [2,51,52] shows best fits with nearly the same $\Delta_0 \sim 4.7$ meV and $\alpha = 1.7$ for $\rho_{ab}$ [2,52] (for $\rho_c$ [2,51,52], $\Delta_0 \sim 3.3$ meV & $\alpha = 1.7$). Note that the INS resonance energy, $E_1(T)$ [14], can be described well with these parameters, suggesting $\Delta_{ab} \sim E_1$. A recent band structure calculation [7] identifies the $\mathbf{Q_1}$ resonance [15] as originating from FS nesting. The association $\Delta_{ab} \sim E_1$ is likely to be valid since the same gapped resistive behavior and $\mathbf{Q_1}$ resonance are seen to extend into the AF phase with both $E_1$ and $\Delta_{ab}$ increasing [17]. These observations also indicate $E_1$ and $\Delta_{ab}$ are not the HO gap. Furthermore, $\Delta_{ab}$ decreases very little with increasing $H$ when $H \parallel I \parallel ab$ [51], while both $T_{HO}$ and $\Delta_c$ decrease but $E_0$ increases when $H \parallel I \parallel c$ [53,54] (cf. Our $\Delta_{hyb}$ remains constant up to 4 T, S6 in [38]). This indicates that $\Delta_c$ (and $\Delta_{ab}$) is not of the same origin as for the $\mathbf{Q_0}$ resonance associated with HO. This field dependence could be explained by a change in the HO-induced nesting vector (thus, decrease in $E_1$) due to decreased $T_{HO}$. Therefore, the resistive gaps are likely to be magnetic in nature and unlikely the HO gap (S5 in [38]). These same magnetic excitations are not distinctly observed in QPS conductance because bosonic excitations such as phonons give weak features requiring more sensitive second harmonic measurements [28]. Our NZBC behavior in Fig. 4c may indicate the effect of magnetic excitations indirectly via interaction with charge carriers [55]. A second harmonic measurement with a current along the $a$- or $b$-axis (i.e., $\parallel \mathbf{Q_1}$) at low $T$ is planned.

The hybridization gap being distinct from the HO parameter is consistent with the general concept that the gradual hybridization process is unlikely to cause a phase transition. Following a generic argument, $T_{hyb}$ (~27 K) may be the temperature below which a coherent heavy Fermi liquid emerges, i.e., $T_{hyb} = T_{coh}$. This association is different from a conventional one, $T_{coh} = T^*$, defined from a resistivity peak.

For URu$_2$Si$_2$ (Fig. 2a), $T^* = 70$-$80$ K $\gg T_{hyb}$. However, $T^*$ may only signify a crossover in the dominant transport scattering channel, whereas $T_{coh}$ is indicative of fully developed coherence among the renormalized Bloch states. Interestingly, our $T_{hyb}$ is close to the temperature for the Fermi liquid behavior ($\propto T^2$) [56], supporting this speculation, but the nature of emergent heavy fermions in URu$_2$Si$_2$ is a topic of continued debate [52,56]. Additionally, the difference between $T_{hyb}$ and $T_{HO}$ is so large that the fluctuating HO scenario [57] may not account for our results.

We now discuss crucial elements to resolving the HO problem. Pressure and magnetic field play quite different roles in URu$_2$Si$_2$: Pressure induces AF order but magnetic field resurrects HO [13]. While both phases exhibit the **Q$_1$** resonance, the **Q$_0$** resonance is unique to the HO, albeit the AF ordering occurs at the same wave vector [16]. Clearly, this points to the crucial roles played by the **Q$_0$** resonance. The recent theory based on band calculations [7] is in discrepancy with our QPS results since the suggested FS gapping along Γ-M should be detected as a dramatic change at $T_{HO}$ in our QPS spectra. Our above analysis suggests the **Q$_1$** resonance may cause the gapped resistive behavior but does not affect the conductance dramatically. Thus, we conjecture the HO, which does not originate from itinerant bands, induces the FS nesting. No multipolar orders predicted to arise from localized *f*-electrons have been detected. Even though crystal field effects are not established, they well deserve a revisit [6,9] to determine what crucial roles are played by the local degrees of freedom for: i) strong uniaxial magnetic anisotropy as observed by neutron scattering [58] and magnetic susceptibility [1]; ii) interplay of pressure and magnetic field in tuning the crystal field *f*-levels and the inter-site interaction [13].

In conclusion, our QPS on URu$_2$Si$_2$ unambiguously detects a novel Fano resonance as predicted for a Kondo lattice and probes the hybridization gap in the renormalized heavy bands. This gap opens at $T_{hyb} \sim 27$ K $\gg T_{HO}$, indicating it is not the HO parameter. Our analysis of the gapped resistivity behavior suggests gapped magnetic excitations rather than a FS gapping as its origin, consistent with no dramatic change in QPS at $T_{HO}$. Further detailed studies as a function of magnetic field and pressure are planned, and expanding our investigation into another Kondo lattice system, UPd$_2$Al$_3$, is of immediate interest due to its strikingly similar properties but with a known AF phase of localized nature [59]. Also, other comparative studies will be fruitful, including intermediate valence vs. Kondo regime or Ce (one *f*-electron) vs. Yb (one *f*-hole) compounds.

We thank H. Arham & C. Lam for their experimental help, P. Chandra, P. Coleman, M. Dzero, P. Ghaemi, C. Hunt, P. Riseborough, and J. Schmalian for fruitful discussions. This material is based upon work supported by the U.S. Department of Energy, Division of Materials Sciences under Award No. DE-FG02-07ER46453, through the Frederick Seitz Materials Research Laboratory at the University of Illinois at Urbana-Champaign. The work at the Los Alamos National Laboratory is carried out under the auspices of the U.S. Department of Energy, Office of Science.

**FIGURE CAPTIONS**

**Figure 1. a**, Single impurity Fano resonance. $T_K$: Kondo temperature. **b,** Hybridization between a conduction band ($\varepsilon_k$) and localized states ($\varepsilon_f$) (see text). $\mu$ is the chemical potential. **c**, DOS for the renormalized heavy bands (thick line), DOS broadened due to correlation effects (dotted line), and $dI/dV$ (thin line) simulating our data at $T < T_{HO}$. **d**, Fano resonance in a Kondo lattice (KL). $T_K$: characteristic temperature for the KL. Inset: schematic for QPS. In the ballistic regime, an incident QP passes through the interface and scatters into the bulk bands.

**Figure 2. a**, $ab$-plane resistivity (resistance) for $URu_2Si_2$ ($U(Ru_{0.985}Rh_{0.015})_2Si_2$). $T^*$ for resistance maximum is 82 K and 70 K, respectively. Not the large RRR ($\equiv R_{300K}/R_{T\to0K}$): 248 for $URu_2Si_2$ and 5.8 for $U(Ru_{0.985}Rh_{0.015})_2Si_2$. $T_{HO}$, taken for minimum in $dR/dT$, is 17.56 K and 12.8 K, respectively. The lines are best fits with gapped AF excitations: $\rho_{AF1}$ (solid red, $\Delta$=4.9 meV)) and $\rho_{AF2}$ (dotted blue, $\Delta$=5.1 meV) (see text). **b**, Specific heat divided by temperature for $URu_2Si_2$.

**Figure 3. a & b**, Differential conductance (normalized by $dI/dV$ at -50 mV) curves for junctions along the $c$-axis of $URu_2Si_2$ and $U(Ru_{0.985}Rh_{0.015})_2Si_2$, respectively. Curves are shifted vertically. Dotted lines are a guide to the eye. **a**, The measurement temperature (the differential junction resistance, $R_J$, at −50 mV) is 3.49 (12.3), 3.51 (18.7), 2.07 (16.7), 4.41 (55.6) and 4.35 K (51.0 Ω) for the curves from 1 to 5, respectively. **b**, The measurement temperature is 4.34 K for all junctions and $R_J$ is 19.5, 25.0, 23.5, 20.4 and 19.7 Ω for the curves from 1 to 5, respectively. **c-f**, Typical conductance spectra for $URu_2Si_2$ and best fit curves with parameters shown in the table.

**Figure 4. a**, Temperature dependent conductance (circles, normalized by $dI/dV$ at -30 mV) and fit curves (lines). Note that reasonably good fits are obtained over the full temperature range. The $R_J$ at the lowest

temperature is 19.1 Ω. The top three curves are plotted on an expanded vertical scale. **b**, Temperature dependence of the hybridization gap (solid circles), $\Delta_{hyb}$, opening at $T_{hyb} \sim 27$ K $\gg T_{HO}$ and the renormalized $f$-level, $\lambda$, (right axis, open circles). **c**, $R_J$ at zero bias (open circles) and at -25 mV (crosses), and the NZBC (right axis, solid circles), indicating the junctions are in the spectroscopic regime.

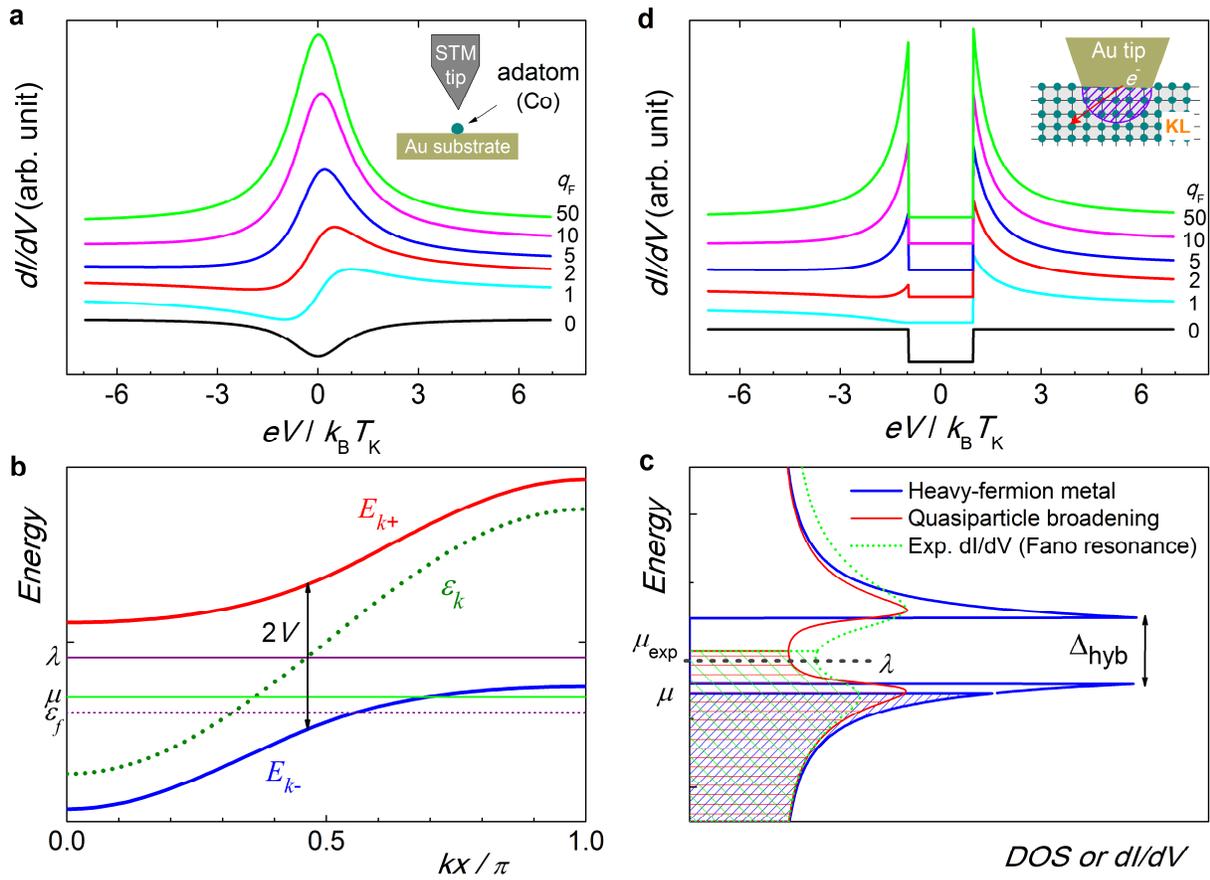

Figure 1. W. K. Park *et al*.

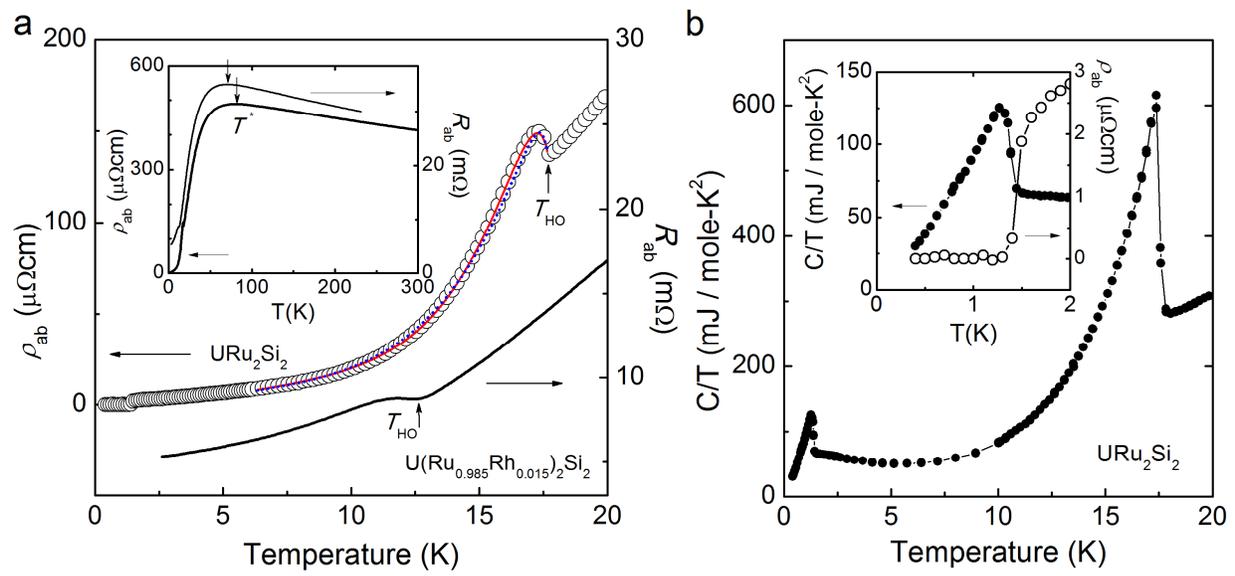

Figure 2. W. K. Park *et al.*

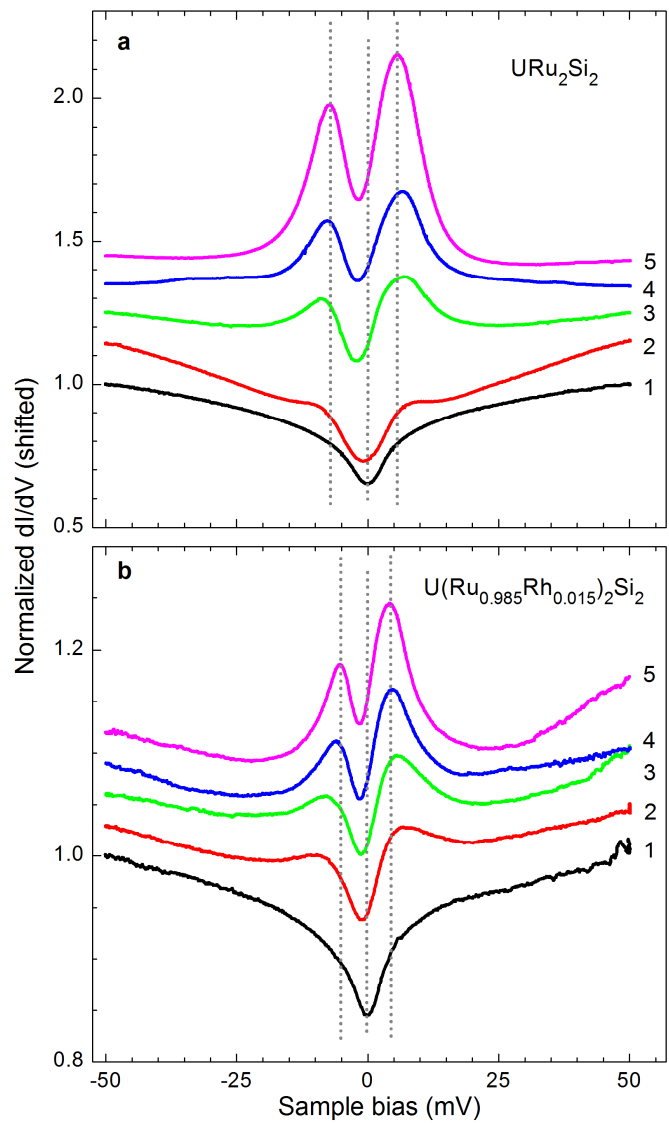
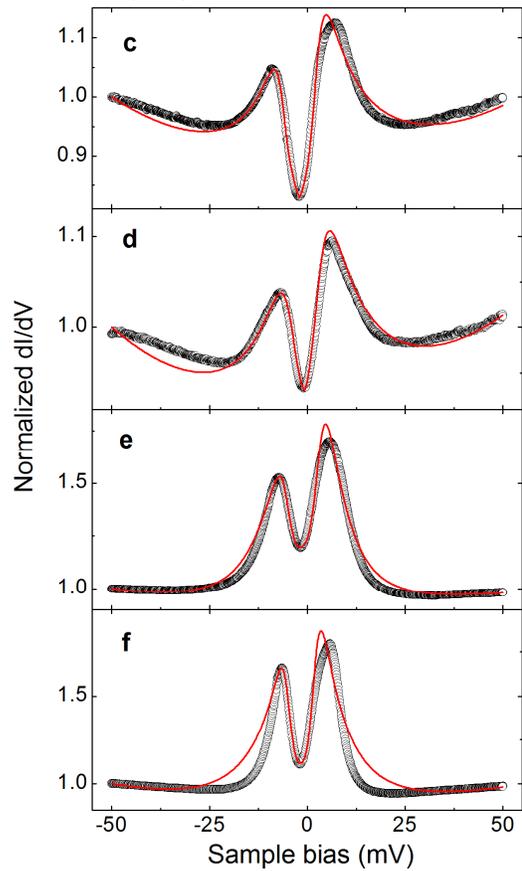

Figure 3. W. K. Park *et al*.

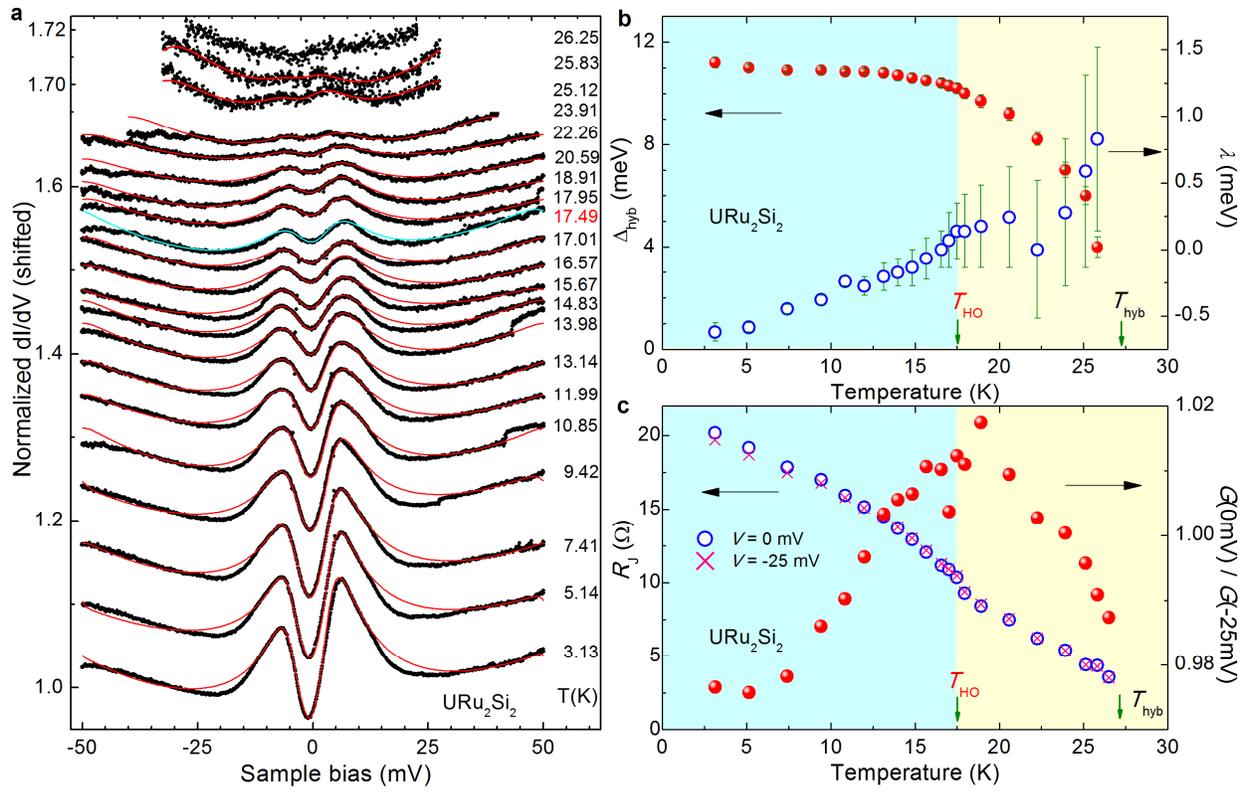

Figure 4. W. K. Park *et al*.

**Online Supplementary Material**

**Observation of the hybridization gap and Fano resonance in the Kondo lattice URu$_2$Si$_2$**

W. K. Park[1], P. H. Tobash[2], F. Ronning[2], E. D. Bauer[2], J. L. Sarrao[2], J. D. Thompson[2], and L. H. Greene[1]

[1]*Department of Physics and the Frederick Seitz Material Research Laboratory, University of Illinois at Urbana-Champaign, Urbana, Illinois 61801, USA*
[2]*Los Alamos National Laboratory, Los Alamos, New Mexico 87545, USA*

**S1. Supporting Evidence for a Fano Resonance**

As demonstrated in Fig. 1d, theoretically, the shape of a conductance curve can vary widely depending on the Fano parameter, $q_F$. This is observed experimentally in our data shown in the main text (Fig. 3) and further demonstrated here in Fig. S1. This series of conductance curves are obtained as the tip is brought into contact with the crystal further and further, as reflected by a change in the junction resistance $R_J$ measured at a negative maximum bias. Here, the conductance shape change can be explained as due to a varying $q_F$ value that is determined by the relative probability for scattering into the heavy electron bands and into the conduction band in URu$_2$Si$_2$ (see Fig. 1d).

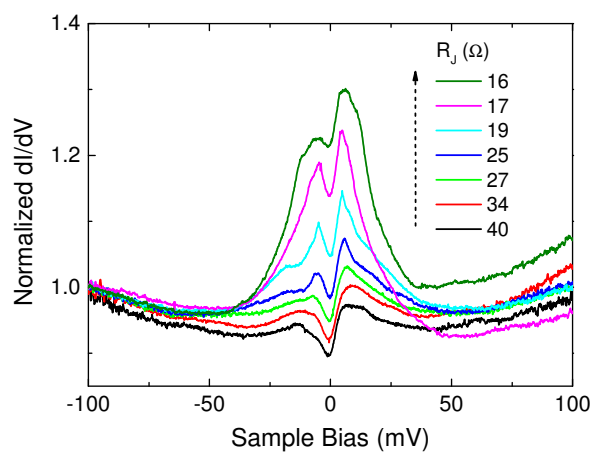

**Fig. S1:** Change in the conductance shape as a function of decreasing $R_J$ as the tip is driven toward the URu$_2$Si$_2$ crystal, implying a change in $q_F$ due to a change in the relative coupling strengths to the heavy fermion and conduction bands. Some additional structures other than the main double-peak structure may arise from a non-trivial junction geometry as the tip is pushed against the crystal surface which may not be atomically smooth over the junction area.

**S2. Details for Data Analysis and Importance of Cleaner Data**

Our model fitting for the conductance data shown in Figs. 3c-f and Fig. 4a was carried out by numerical computation using a MATLAB code. All involved energies are scaled with respect to a hypothetical Kondo lattice energy scale $T_K = 80k_B$, which corresponds to $T^*$ observed in the resistivity data (Fig. 2a). The fitting was optimized to account for mainly the conductance shape around the double peaks with the background shape adjusted by a parabolic function. We assume symmetric conduction band edges, $D_1 = D_2$, and set $D = 41 T_K$ and $W = T_K$. We use the quasiparticle broadening parameter in the form: $\gamma(E) = c \cdot E^n T_K^{1-n} + \gamma_0$ for $E = eV < d \cdot T_K$ and $\gamma(E) = E/[1+\log(E/(d \cdot T_K))]^2 + \gamma_0$ for $E = eV \geq d \cdot T_K$, where typical parameter values are $c \sim 0.65$, $d \sim 20$, n = 1 - 2, and $\gamma_0$ = 0 - 3 meV. For the temperature dependent data in Fig. 4a, we find best fit curves by convolving the calculated conductance for zero temperature to measurement temperatures using the fast Fourier transform (FFT) technique. The effect of thermal smearing becomes bigger at high temperature, so the nominal peak-to-peak distance deviates from $\Delta_{hyb}$ obtained by fitting the data. The error bars for $\Delta_{hyb}$ and $\lambda$ in Fig. 4b were estimated with a criterion that the standard deviation of calculated conductance from the experimental data over the bias region covering twice the hybridization gap around zero bias doesn't exceed ~5% of that for an optimal fit curve.

A better fit could be obtained by using data-specific background functions, which are not necessarily parabolic, but we do not perform such an extensive fitting since our current fitting produces satisfactorily matching curves, enabling us to extract the hybridization gap in a consistent way. This is possible because most of our analyzed data exhibit sharp peak structures, in which case we also find that for lowest measurement temperatures the peak-to-peak distance in the raw data is close to that obtained from a curve fitting. Note that our characteristic conductance features are among the sharpest reported in the literature [36,37,60-66] (see below, S3). The distinct asymmetric double-peak structure observed reproducibly in our measurements is a crucial aspect that leads us to conjecture on a Fano resonance [23] as their physical origin and to determine the hybridization gap opening at a temperature much higher than $T_{HO}$. The latter is not possible if the conductance curves are broadened too much (see below, S3 & S4). Moreover, this ballistic nature of the junctions we measured allows us to observe the ZBC behavior shown in Fig. 4c that does not simply follow the bulk resistivity. As is known, the stronger the similarity between $R_J(T)$ and $\rho(T)$ is, the conductance data are less likely to contain spectroscopic information.

## S3. Supporting Evidence for the Hybridization Gap Opening at $T_{hyb} \gg T_{HO}$

The split peaks reflecting the characteristic DOS due to hybridization are frequently observed well above $T_{HO}$. In addition to the temperature-dependent conductance spectra presented in Fig. 4a, here we report another set of data in Fig. S2. Although a full set of temperature dependent data are not obtained, it is clear that the peaks merge at a temperature between 24 K and 30 K, similarly to the data presented in Fig. 4a. The hybridization gap and the renormalized *f*-level extracted from the Fano resonance model fit are shown in Fig. S2b. Qualitatively, they are similar to those reported in Fig. 4b.

In the literature [36,37,60-66], a range of different gap opening temperatures have been reported, albeit most of them are higher than $T_{HO}$. As reconciliation for this behavior, it was suggested [65] that the pressure exerted by a tip might increase $T_{HO}$ locally. However, to account for our $T_{hyb}$ being higher than $T_{HO}$ by almost 10 K, the local pressure should be on the order of several tens of GPa, an extreme scenario. Here, we provide more plausible explanations. We note at high temperature the hybridization-gap peaks can appear to merge faster than the intrinsic gap depending on the extent of smearing (both quasiparticle and thermal). For this reason, it is important to obtain conductance spectra showing distinctly split peaks at low temperature in order to extract intrinsic temperature dependence of the gap.

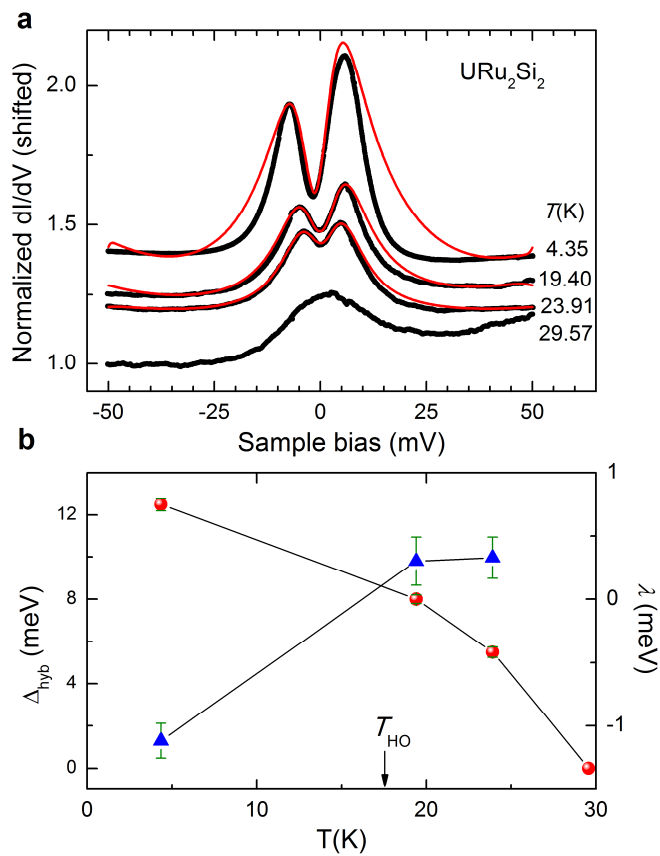

**Fig. S2: a,** Temperature dependent conductance spectra in $URu_2Si_2$. Filled circles are data and red solid lines are fits to the data using the Fano resonance model (see main text). There were changes in the junction characteristics during warm-up. **b**, Temperature dependences of the hybridization gap ($\Delta_{hyb}$, left axis, filled red circles) and the renormalized $f$-level ($\lambda$, right axis, filled blue triangles) extracted from the analysis. Lines are a guide to the eye. It is clearly seen that the gap persists well above $T_{HO}$, closing at a temperature between 24 K and 30 K. This behavior, along with that for $\lambda$, is qualitatively similar to that in Fig. 4b.

To demonstrate this point more explicitly, we analyze the data reported in the literature [36,37,60-66] and plot the reported gap opening temperature ($T_{gap}$) vs. the sharpness of the reported low-temperature conductance curve in Fig. S3. Here, we quantify the sharpness as the ratio of the peak conductance (maximum) at a positive bias to the dip conductance (minimum) at a negative bias. The error bars are inferred from the conductance shape at the highest measurement temperature. Apparently, there is a strong positive correlation between $T_{gap}$ and the sharpness. The three works [60,61,66] marked on the lower left (in red diamonds) report $T_{gap}$ to be close to $T_{HO}$. We note that their $G(V)$ curves exhibit additional structures (peaks or humps) outside the main peaks due to the hybridization gap. Their peak positions are not only symmetric with respect to zero bias, unlike the main hybridization peaks, but also change from junction to junction, probably ruling out a spectroscopic origin such as FS gapping [60,61,66]. Indeed, our analysis (not shown) shows that those additional peak positions fit to the local heating model [28,67] quite well, indicating they may originate from the junction being thermal (increasing bias dissipates more heat in the junction area, driving the junction toward the thermal limit). Also, the bias dependence of $R_J$ is found to fit to the bulk resistivity using the model adopted in the main text, with temperature replaced by the bias voltage, again implying the thermal nature of the junctions. On the contrary, the three works [36,62] marked on the upper right in Fig. S3, including ours, report higher $T_{gap}$ values and enhanced sharpness from the data that do not show any such additional peaks, consistent with our argument on the spectroscopic nature.

Similarly contrasting behaviors are also evidenced in the zero-bias junction resistance ($R_0$) vs. temperature data. The same three papers [60,61,66] on the lower left in Fig. S3 report $R_0$ data closely following the bulk resistivity. It has been argued this observation confirms QPS measuring the bulk property. However, it is known that the strong similarity between $R_0(T)$ and $\rho(T)$ suggests that the junction is closer to the thermal limit [67]. This is exactly how QPS differs from a bulk conductivity measurement. In the latter, no spectroscopic information such as the gap size can be derived. On the other hand, opposite temperature dependences of $R_0$ are observed in the other reports [36,62] marked on the upper right in Fig. S3, including ours, and also in [37]: it increases with decreasing temperature, opposite to the bulk resistivity. This is because, here in the spectroscopic regime, $R_0$ is governed by the energy-resolved bulk electronic properties including the DOS and the hybridization gap. Therefore, we argue that the gap opening near $T_{HO}$ reported in [60,61,66] may be an artifact due to local heating or large smearing effects. To determine $T_{gap}$ accurately, it is crucial to obtain a junction showing the hybridization-gap peaks as sharp as possible at low temperature, as demonstrated in our work reported in this paper.

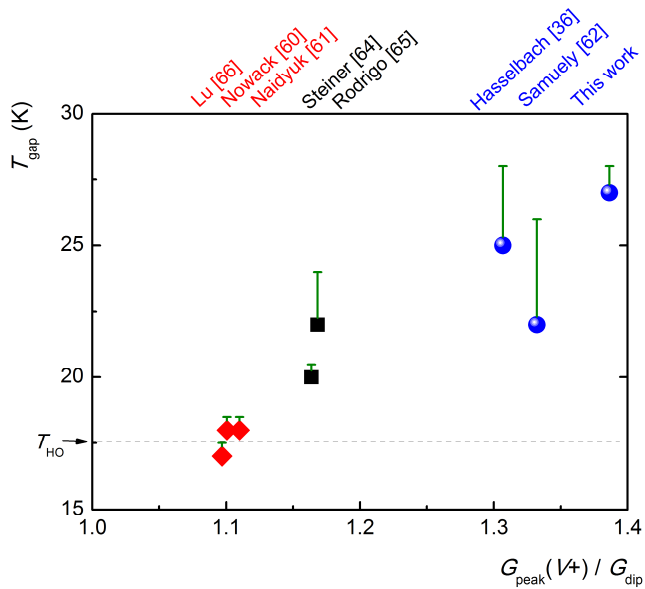

**Fig. S3:** Comparison of gap opening temperature ($T_{gap}$) vs. sharpness of the low-temperature gap structure in the conductance data reported by different groups. The sharpness is quantified as the ratio of the positive-peak conductance (maximum) to the dip conductance (minimum). The error bars are inferred from the conductance shape at the highest measurement temperature. See text for the discussion of the differences between the lower left (red diamonds) and the upper right (blue circles) groups.

## S4. Effect of Quasiparticle Broadening on the Hybridization-Gap Peak Structure

As already shown in the original paper by Coleman and coworkers [23], a large quasiparticle broadening parameter, $\gamma$, smears out an otherwise distinct double-peak structure. Our conductance spectra presented in the paper exhibit clear split peaks, indicating that the quasiparticle broadening effect is not dominant to smear them out completely to produce a single impurity Fano line shape [26]. In order to visualize this effect more quantitatively, we simulate conductance curves for two different values of $q_F$, 2 and 10, as a function of $\gamma$. As shown in Fig. S4, the quasiparticle broadening effect renders the double peak structure to be smeared out more quickly for a smaller $q_F$ value. As discussed in the main text, the STM data [21,22] exhibit much smaller $q_F$ values (< 2) compared to ours ($q_F \sim 10$). Thus, one may speculate that weaker coupling to the heavy fermion bands in the STM configuration may cause the hybridization peak structure to be smeared out more easily, resulting in a single impurity line shape as argued by Wölfle and coworkers [26]. Then, the question is how the gap-like feature is detected in the STM measurements if the broadening effect is that large. As an alternative scenario, we discuss a possibility of reduced hybridization at the surface due to smaller near-neighbor coordination, as presented in the main text. In any case, our measurement is less sensitive to the surface and couples more strongly to the hybridized heavy bands (large $q_F$) so that the double-peak structure survives under the quasiparticle broadening effects.

## S5. Effects of Magnetic Excitations on the Hall Effect

The same reasoning as adopted for the analysis of the gapped resistive behavior may apply to the Hall effect [68,69] since it also measures scattering of charge carriers. The Hall coefficient in $URu_2Si_2$ jumps at $T_{HO}$ and then decays weakly with decreasing temperature. This is not explainable solely by a simple carrier depletion picture or Fermi surface gapping, in which Hall coefficient is expected to remain constant or to keep increasing after the initial jump at the transition as observed in some known antiferromagnetic compounds including $Ba(Fe,Co)_2As_2$ [70] and $U_2Zn_{17}$ [71]. Therefore, it may be appropriate to account for the unusual behavior of the Hall effect in $URu_2Si_2$ by including the contribution from scattering off gapped magnetic excitations though keeping in mind the nearly compensated electronic nature of $URu_2Si_2$.

## S6. Magnetic Field Dependence

Theoretically, magnetic field-induced Zeeman effect can split a spin-degenerate band, possibly changing the DOS. This could be explored as a test for the hybridization model, as hypothesized to explain experimental results on Kondo insulators [72]. Figure S5 shows conductance curves taken under magnetic field applied along the *c*-axis of $URu_2Si_2$. The peak positions remain almost the same up to 4 Tesla, whereas the peak intensity increases with field, unlike superconductive tunneling or Andreev reflection in which magnetic field typically broadens the peak structure. The insensitiveness of the peak position, or $\Delta_{hyb}$, to magnetic field is in contrast to the resistive behavior (see main text) for both field and current along the *c* axis, supporting that the hybridization gap is different in nature from the resistive gap. For future investigations, it may be worthwhile to make measurements under much stronger fields [73], which may cause noticeable changes in the hybridized bands and the gap.

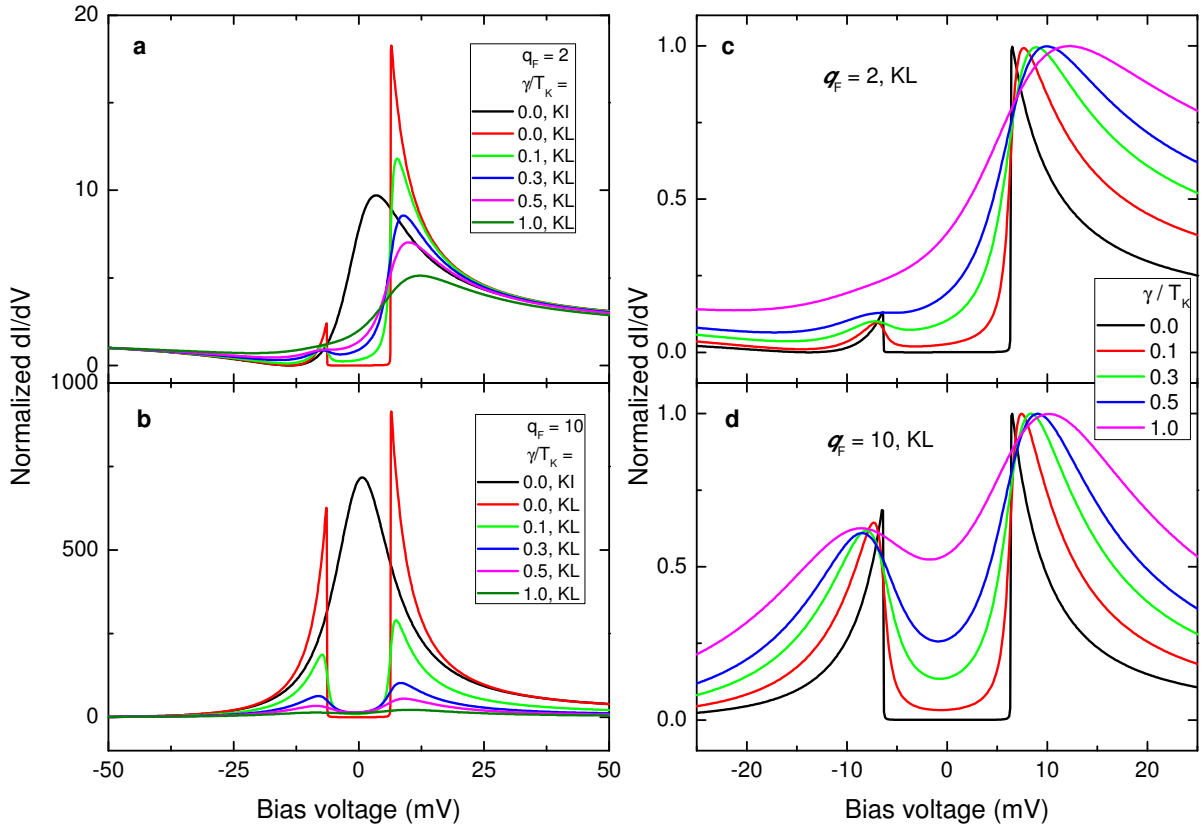

**Fig. S4**: Effect of quasiparticle lifetime broadening ($\gamma$). **a & b,** Normalized conductance curves calculated for $q_F$ =2 and 10, respectively, with $\Delta_{hyb}$ = 13 meV and constant $\gamma$ as indicated in the figure. Also shown are single impurity line shapes for comparison. **c & d,** Expanded views by normalizing to the peak on the positive bias side. The double peak structure is smeared rapidly for $q_F$=2, whereas it is visible up to a higher $\gamma$ value for $q_F$=10.

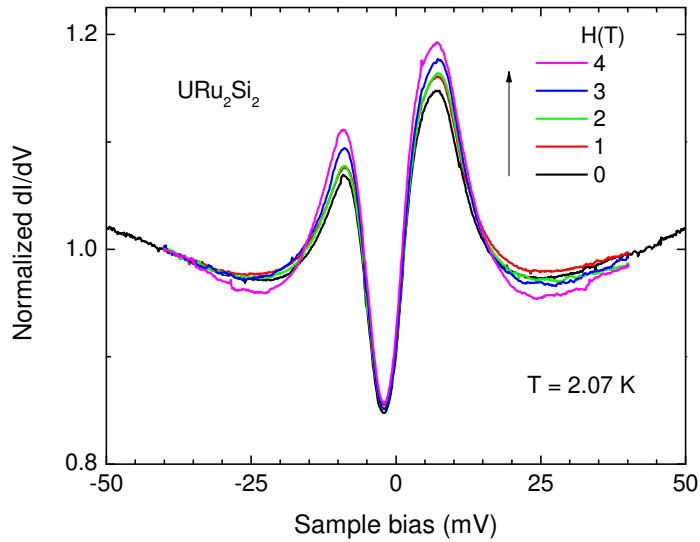

**Fig. S5:** Magnetic field dependence of the differential conductance in URu$_2$Si$_2$. The field is applied along the *c*-axis. The junction resistance remains in the range of 16.5 – 17.6 Ω.